# Time-dependent Bragg diffraction bymultilayer gratings

Jean-Michel André[1,2]*, Philippe Jonnard[1,2]

[1] *CNRS UMR 7614, Laboratoire de Chimie Physique - Matière et Rayonnement, 11 rue Pierre et Marie Curie, F-75231 Paris cedex 05, France*

[2] *Sorbonne Universités, UPMC Univ Paris 06, Laboratoire de Chimie Physique - Matière et Rayonnement, 11 rue Pierre et Marie Curie, F-75231 Paris cedex 05, France*

*corresponding author : jean-michel.andre1@upmc.fr

**Abstract**

The time-dependent Bragg diffraction by multilayer gratings working by reflection or by transmission is investigated. The study is performed by generalizing the time-dependent coupled-wave theory previously developed for one-dimensional photonic crystal [André J-M and Jonnard P, J. Opt. 17, 085609 (2015)] and also by extending the Takagi-Taupin approach of the dynamical theory of diffraction. The indicial response is calculated. It presents a time-delay with a transient time that is a function of the extinction length for reflection geometry and of the extinction length combined with the thickness of the grating for transmission geometry.

# 1. Introduction

The time-dependentBragg diffraction of a one-dimensional photonic crystal (1D-PC) modelled by a periodic stack of bilayers has been studied in our recent paper [1]. In the present work, we extend our analysis to multilayer gratings that can be regarded as a specific case of two-dimensional photonic crystals. This kind of multilayer gratings has acquired a considerable importance in optics of short wavelength radiation (from UV to hard x-rays) [2–5]. The advent of short (femtosecond) and even ultra-short (attosecond) sources in these spectral domains has naturally led to a growing interest in the study of the temporal response of these optics. In fact, most of the works relative to the temporal diffraction have concerned the "real" crystals usually implemented in x-ray diffractive optics. Chukovskii and Förster have found analytical solutions of the time-dependent Bragg diffraction by thick crystal [6] using a Tagaki-Taupin approach; Graeff has found similar solutions but for Laue geometry [7]. More recently the problem has been treated in reflection geometry and transmission geometry by Lindberg and Shvyd'ko,also within the Tagaki-Taupin approach, with application to self-seeding free-electron lasers [8,9]. Fourier analysis, which is not very adequate to treat transient phenomena, has been used by many authors to investigate spatiotemporal response of crystals or multilayer optics [10,11].

In this paper, we implement the time-dependentcoupled-wave theory (CWT) in the two-wave approximation,which leads to a system of coupled partial differential equations (PDEs); note that this CWT has been recently unified in [12] to treat the diffraction by multilayer gratings in the steady-state case.For the time-dependent case,we solve this system using the matrix methoddeveloped in [1]but we also use the Tagaki-Taupin approach (in quasi-specular conditions) to give a physical insight to the problemand to point out the key physical parameters of the problem.We consider multilayer gratings with various shapes (lamellar, sliced, blazed,…)diffracting in reflectionor transmission geometry. Werestrict our analysis to *s*-polarization (transverse electric) case and planar (non-conical) diffraction since the other situations, *p*-polarization and conical diffraction,are more complicated to handle because the paraxial approximation cannot be directly applied.

# 2. Time-dependent coupled-wave theory

We consider a diffraction multilayer grating as shown in Figure 1 struck by an incident plane wave under a glancing angle $\theta_0$ with a wave-vector $\hat{k} = \cos\theta_0\,\hat{x} + \sin\theta_0\,\hat{z}$ forming together with the normal to the grating surface, the incidence plane (x, z). The diffraction is considered as planar (not conical). Let us recall that in conical mounting the incident wave-vector is not perpendicular to the grating grooves and the wave-vectors of the different diffraction orders lie on a conical surface; in standard planar mounting that interests us in this work, the wave-vectors of the different diffracted orders remain in the incident plane. The considered geometry of the grating covers a large set of grating shapes, from the simple ($\Gamma = 1, \phi = 0$) or sliced multilayer mirror ($\Gamma = 1, \phi \neq 0$) to the lamellar multilayer grating ($\Gamma < 1, \phi = 0$) or blazed multilayer grating ($\Gamma < 1, \phi \neq 0$), were $\Gamma$ is the ratio of the pitch to the grating period and $\phi$ defines the orientation of the multilayer with respect to the surface.

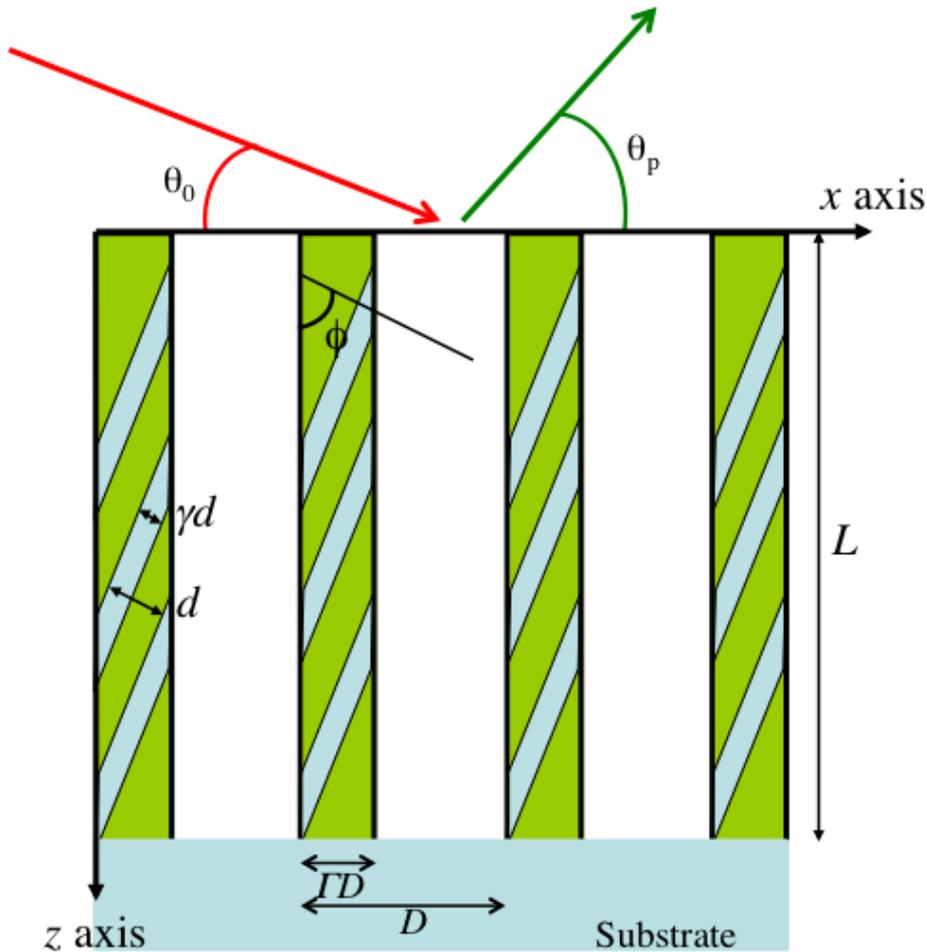

*Figure 1: Sketch of the grating of period D; each grating pitch has a width equal to ΓD and is formed by a periodic stack of N bilayers with thickness d; the bilayer is made up of a material **a** of dielectric susceptibility $\chi_a$ and material **b** of dielectric susceptibility $\chi_b$, with layer thickness $d_a = \gamma d$ and $d_b = (1-\gamma)d$ respectively. The incoming radiation strikes the*

*multilayer structure under a glancing angle $\theta_0$ with a wave-vector $\hat{k}$ in the $(x, z)$ plane. A Cartesian orthogonal reference frame $(\hat{x}, \hat{y}, \hat{z})$ is used.*

Since the propagation takes place within the $(x, z)$ plane then, in the framework of the paraxial approximation, the wave time-dependent propagation equation is given in Gaussian cgs units by

$$\frac{\partial^2 E(R_\parallel, t)}{\partial x^2} + \frac{\partial^2 E(R_\parallel, t)}{\partial z^2} - \frac{1}{c^2}\frac{\partial^2 E(R_\parallel, t)}{\partial t^2} = \frac{4\pi}{c^2}\frac{\partial^2 P(R_\parallel, t)}{\partial t^2}$$

(1)

where $E(R_\parallel, t)$ stands for the electric field and $P(R_\parallel, t)$ for the electric polarization vector, $c$ being the speed of light in vacuum and $R_\parallel = (x, z)$ the position vector in the $(x, z)$ plane. Let us recall that the paraxial approximation neglects the term $\nabla(\nabla \cdot E)$ in the Maxwell equations leading to Eq.(1); this approximation is justified as long as the electric field remains transverse.

Since we consider only the *s*-polarization (transverse electric) case, the electric field vector $E(R_\parallel, t)$ is along the *y* axis. It is assumed that the electric field of the optical pulse is formed of a quickly varying carrier with frequency $\omega$, ⍰ ⍰ ⍰ ⍰ ⍰ ⍰ ⍰ ⍰ ⍰ by an envelope $E_0(R_\parallel, t)$ and we write it as follows taking into account the *s*-polarization geometry

$$E(R_\parallel, t) = E_0(R_\parallel, t) e^{-i\omega t} \hat{y}$$

(2)

We assume that the polarization is essentially electronic and follows instantly the change of the electric field, and that the media have a linear response. Hence we write the polarization $P$ as

$$P(R_\parallel, t) = \chi(R_\parallel) E(R_\parallel, t) = \chi(R_\parallel) E_0(R_\parallel, t) e^{-i\omega t} \hat{y}$$

(3)

where $\chi(R_\parallel, z)$ is the dielectric susceptibility assumed in our model to be time independent.

The susceptibility in the grating can be expanded in Fourier series[12]

$$\chi(R_\parallel) = \sum_{m,n=-\infty}^{+\infty} u_m U_{m,n} e^{i G_{mn} \cdot R_\parallel}$$

(4a)

$$G_{mn} = m G_x \hat{x} + n G_z \hat{z}$$

(4b)

$$G_x = \frac{2\pi}{D} \sin\phi \; ; G_z = \frac{2\pi}{d} \cos\phi$$

(4c)

$$u_0 = \bar{\chi} = \chi_a \gamma + \chi_b (1-\gamma)$$

(4d)

$$u_m = \frac{-i}{2 m \pi} \Delta\chi \left(1 - e^{-2i\pi m \gamma}\right) ; \Delta\chi = \chi_a - \chi_b$$

(4e)

$$U_{m,n} = \Gamma \, sinc\left[\Gamma\left(m \frac{D \sin\phi}{d} - n\right)\right]$$

(4f)

Inserting Eqs.(3,4) in Eq.(2) leads to the following equation for the spatiotemporal propagation of the electric field envelope

$$\frac{\partial^2 E_0(\mathbf{R}_\parallel, t)}{\partial x^2} + \frac{\partial^2 E_0(\mathbf{R}_\parallel, t)}{\partial z^2} + k^2 \, \epsilon(\mathbf{R}_\parallel) E_0(\mathbf{R}_\parallel, t) + 2i \frac{\omega}{c^2} \epsilon(\mathbf{R}_\parallel) \frac{\partial E_0(\mathbf{R}_\parallel, t)}{\partial t}$$
$$- \frac{\epsilon(\mathbf{R}_\parallel)}{c^2} \frac{\partial^2 E_0(\mathbf{R}_\parallel, t)}{\partial t^2} = 0$$

(5)

with $\epsilon$ the dielectric constant given in the Gauss unit system, by

$$\epsilon(\mathbf{R}_\parallel) = 1 + 4\pi \, \chi(\mathbf{R}_\parallel)$$

(6)

and

$$k = \frac{\omega}{c}$$

(7)

The envelope of the diffracted electric field can be represented by the Rayleigh expansion [13]

$$E_0(\mathbf{R}_\parallel, t) = \sum_{p=-\infty}^{+\infty} \mathcal{E}_p(z, t) \exp(i \, q_p \, x)$$

(8)

with

$$q_0 = k \cos\theta_0$$

(9a)

$$q_p = q_0 + p \, G_x \; ; p \neq 0$$

(9b)

At this step we apply the slowly varying amplitude approximation (SVA) in time: the second derivatives with respect to time are neglected. As a result, Eq.(5) can be reduced to a time-dependent Schrödinger equation for each field coefficient $\mathcal{E}_p(z,t)$ of the Rayleigh expansion

$$\left(\frac{\partial^2}{\partial z^2} + \kappa_p(\boldsymbol{R}_\parallel)^2 + 2\,i\,\epsilon(\boldsymbol{R}_\parallel)\frac{\omega}{c^2}\frac{\partial}{\partial t}\right)\mathcal{E}_p(z,t) = 0$$

(10a)

$$\kappa_p(\boldsymbol{R}_\parallel)^2 = k^2\epsilon(\boldsymbol{R}_\parallel) - q_p^2$$

(10b)

Note that at this step, the SVA in space has not yet been applied. Eqs.(10) could be treated by means of numerical codes developed to solve the time-dependent Schrödinger equation. We choose in this work to use instead the approach of CWT which proved to be very efficient in this kind of optical problem[14]. In this context, the field can be written as the superposition of two waves propagating in opposite directions along the z-axis, so that we write, using the CWT

$$\mathcal{E}_p(z,t) = F_p(z,t)e^{+i\overline{\kappa_p}z} + B_p(z,t)e^{-i\overline{\kappa_p}z}\ ;\ \overline{\kappa_p} = \sqrt{k^2\bar{\epsilon} - q_p^2}$$

(11a)

together with the following requirement to ensure the uniqueness [12]

$$F_p^{'}(z,t)e^{+i\overline{\kappa_p}z} + B_p^{'}(z,t)e^{-i\overline{\kappa_p}z} = 0$$

(11b)

The term $\bar{\epsilon}$ is the average value of the dielectric constant (term zero of the Fourier series) that means that $\overline{\kappa_p}$ does not depend on the z variable in agreement with the essence of the CWT.

Substituting Eqs.(11) in Eqs.(10) and applying the SVA in space leads to a system of coupled PDEs for each component of the $p^{th}$ term of the Rayleigh expansion

$$\frac{\partial}{\partial z}F_p(z,t) = -\frac{k^2}{2\,\overline{\kappa_p}}\sum_{j,n=-\infty}^{+\infty} u_j U_{j,p-n}\,exp(i\,j\,G_z\,z)\big(F_n(z,t)e^{i(\overline{\kappa_n}-\overline{\kappa_p})z}$$

$$+ B_n(z,t)e^{-i(\overline{\kappa_n}+\overline{\kappa_p})z}\big) - 2\,i\,\bar{\epsilon}\frac{\omega}{\overline{\kappa_p}\,c^2}\frac{\partial}{\partial t}F_p(z,t)$$

(12a)

$$\frac{\partial}{\partial z} B_p(z,t) = \frac{k^2}{2\,\overline{\kappa_p}} \sum_{j,n=-\infty}^{+\infty} u_{-j} U_{-j,p-n}\, exp(-i j\, G_z\, z)\left(F_n(z,t)e^{i(\overline{\kappa_n}+\overline{\kappa_p})z}\right.$$
$$\left. + B_n(z,t)e^{-i(\overline{\kappa_n}-\overline{\kappa_p})z}\right) + 2\,i\,\bar{\epsilon}\frac{\omega}{\overline{\kappa_p}\,c^2}\frac{\partial}{\partial t}B_p(z,t)$$

(12b)

Only the zeroth order term $\bar{\epsilon}$ in the Fourier expansion of $\epsilon(\boldsymbol{R}_\parallel)$ has been kept in the time-dependent term of Eqs.(12); so one can say that our calculation is a first-order (in terms of $\epsilon$) perturbative time-dependent model.

## 3. Two-wave theory

### 3.1. Matrix approach

Let us assume a strong coupling occurring between the incident wave with amplitude $F_0(z,t)$ and the $p^{th}$ wave diffracted by the grating with amplitude $B_p(z,t)$. In this condition it is possible to keep in the system given by Eqs.(12) only the two terms $F_0(z,t)$ and $B_p(z,t)$. This is a situation that corresponds to the so-called two-wave theory (TWT) in the dynamical theory of diffraction. The conditions of validity of this approximation are discussed in several papers [12,15] ; as mentioned in [12] a validity condition for this regime is that the angular width of the diffracted peak is small compared to the distance in terms of glancing angle between the neighbouring diffraction peaks, a condition generally required for a spectroscopic application of the gratings.

One finally gets a system of PDEs with terms that do not depend on the variable $z$, provided than one introduces the following auxiliary amplitude terms

$$f_0(z,t) = F_0(z,t)\, exp[i\left((\overline{\kappa_0}+\overline{\kappa_p})-\frac{j\,G_z}{2}\right)z\,]$$

(13a)

and

$$b_p(z,t) = B_p(z,t)\, exp[-i\left((\overline{\kappa_0}+\overline{\kappa_p})-\frac{j\,G_z}{2}\right)z\,]$$

(13b)

Moreover, if the system is in the vicinity of the *jth* Bragg resonance

$$j\,G_z \approx \overline{\kappa_0}+\overline{\kappa_p}$$

(14)

and $f_0(z,t) \approx F_0(z,t), b_p(z,t) \approx B_p(z,t)$. Eq.(14) gives the generalized Bragg condition. Combining Eqs.(12,13), there results after some algebra the following system of time-dependent coupled PDEs satisfied by the varying amplitudes $f_0(z,t)$ and $b_p(z,t)$ in the domain of the $p^{th}$ diffraction; using the matrix formalism, this system reads

$$\frac{\partial}{\partial z}\overline{\Sigma_B}(z,t) = \overline{\mathcal{B}_{Bp}}\frac{\partial}{\partial t}\overline{\Sigma_B}(z,t) + i\,\overline{\mathcal{M}_{Bp}}\overline{\Sigma_B}(z,t)$$

(15)

where $\overline{\Sigma_B}(z,t)$ is the following column amplitude vector

$$\overline{\Sigma_B}(z,t) = \begin{pmatrix} f_0(z,t) \\ b_p(z,t) \end{pmatrix}$$

(16)

The subscript B stands for the Bragg-case reflection geometry. In Eq.(15), $\overline{\mathcal{M}_{Bp}}$ is the propagation matrix in space

$$\overline{\mathcal{M}_{Bp}} = \begin{pmatrix} -\alpha_B & K_B^+ \\ K_B^- & \tilde{\alpha}_B \end{pmatrix}$$

(17a)

$$\alpha_B = \overline{\kappa} - \frac{j\,G_z}{2} - \frac{k^2}{2\overline{\kappa_0}}\Gamma\,\overline{\chi}\quad ;\overline{\kappa} = \frac{(\overline{\kappa_0} + \overline{\kappa_p})}{2}$$

(17b)

$$\tilde{\alpha}_B = \overline{\kappa} - \frac{j\,G_z}{2} - \frac{k^2}{2\,\overline{\kappa_p}}\Gamma\,\overline{\chi}$$

(17c)

$$K_B^+ = -\frac{k^2}{2\,\overline{\kappa_0}}u_j\,U_{j,-p};\ K_B^- = +\frac{k^2}{2\,\overline{\kappa_p}}u_{-j}\,U_{-j,p}$$

(17d)

$\overline{\mathcal{B}_{Bp}}$ is the propagation matrix in time

$$\overline{\mathcal{B}_{Bp}} = \begin{pmatrix} -\frac{k}{c\overline{\kappa_0}} & 0 \\ 0 & +\frac{k}{c\,\overline{\kappa_p}} \end{pmatrix}$$

(18)

We first consider the time-independent case

$$\frac{\partial}{\partial z}\overline{\Sigma_B}(z) = i\,\overline{\mathcal{M}_{Bp}}\overline{\Sigma_B}(z)$$

(19)

The solution can be obtained by substituting

$$\overline{\Sigma_B}(z) = \begin{pmatrix} A \\ B \end{pmatrix} e^{i\psi z}$$

(20)

and as shown in Appendix 1,

$$\overline{\Sigma_B}(z) = \overline{S_{Bp}}(z)\overline{\Sigma_B}(0)$$

(21)

with

$$\overline{S_{Bp}}(z) = \begin{pmatrix} \dfrac{\exp\left[i\frac{\alpha^- - q}{2}z\right](\alpha^+ + q) - \exp\left[i\frac{\alpha^- + q}{2}z\right](\alpha^+ - q)}{2q} & \dfrac{K^+\left(\exp\left[i\frac{\alpha^- + q}{2}z\right] - \exp\left[i\frac{\alpha^- - q}{2}z\right]\right)}{q} \\ \dfrac{K^-\left(\exp\left[i\frac{\alpha^- + q}{2}z\right] - \exp\left[i\frac{\alpha^- - q}{2}z\right]\right)}{q} & \dfrac{\exp\left[i\frac{\alpha^- + q}{2}z\right](\alpha^+ + q) - \exp\left[i\frac{\alpha^- - q}{2}z\right](\alpha^+ - q)}{2q} \end{pmatrix}$$

(22)

and

$$q = \sqrt{4 K_B^+ K_B^- + \alpha^{+2}}$$

(23)

$$\alpha^+ = \tilde{\alpha}_B + \alpha_B \; ; \; \alpha^- = \tilde{\alpha}_B - \alpha_B$$

(24)

Some calculations show that $\overline{S_{Bp}}(z)$ reduces to the matrix $\bar{S}(z)$ given by Eq.(17) of Ref. [1] for the case of a 1D-PC. From Eqs.(21-24), which form the basis of the time-independent two-wave CW analysis, it is possible to calculate the reflection and transmission diffraction efficiencies of a grating at a given diffraction order and the electric field distribution within the grating.

As for the time-dependent case, one searches the solution by analogy with the time-independent case in the following form:

$$\overline{\Sigma_B}(z,t) = \begin{pmatrix} A(t) \\ B(t) \end{pmatrix} e^{i\psi z}$$

(25)

This approach can be regarded as a kind of Lagrange's method of variation of constants: $A \rightarrow A(t), B \rightarrow B(t)$. Inserting Eq.(25) in Eq.(15) gives after derivation with respect to space

$$\frac{\partial}{\partial t}\begin{pmatrix} A(t) \\ B(t) \end{pmatrix} = -\overline{G_{Bp}}\begin{pmatrix} A(t) \\ B(t) \end{pmatrix}$$

(26)

with

$$\overline{G_{Bp}} = i\,\overline{\mathcal{B}_{Bp}}^{-1}\left(\psi\,\overline{I} - \overline{\mathcal{M}_{Bp}}\right)$$

(27)

Integration of Eq.(25) with respect to time gives

$$\begin{pmatrix} A(t) \\ B(t) \end{pmatrix} = \exp\left(-\overline{G_{Bp}}\,t\right)\begin{pmatrix} A(0) \\ B(0) \end{pmatrix}$$

(28)

Finally, by following a way similar to the one presented for the time-independent case, we obtain

$$\overline{\Sigma_B}(z,t) = \overline{R_{Bp}}(z,t)\overline{\Sigma_B}(0,0)$$

(29)

where

$$\overline{R_{Bp}}(z,t) = \exp\left(-\overline{G_{Bp}}\,t\right)\overline{S_{Bp}}(z)$$

(30)

The propagation of the electric field in time and space (field distribution within the grating, transmission, reflection, …) can be deduced from Eqs.(28-30). In the following, we consider the indicial response in terms of reflection and transmission under Heaviside unit-step input. Some numerical examples will be given in Section 4.

### 3.2 Indicial response in the two-wave approximation

The reflection and transmission coefficients are derived from the initial and boundary conditions: at $z = 0$, a Heaviside unit-step $\Theta(t)$ is applied, so that $f_0(0,t) = \Theta(t)$, and at $z = L$ there is no incoming wave, so that $b_p(L,t) = 0$, which gives from Eq.(29)

$$\begin{pmatrix} f_0(L,t) \\ b_p(L,t) = 0 \end{pmatrix} = \begin{pmatrix} R_{Bp11}(L,t) & R_{Bp12}(L,t) \\ R_{Bp21}(L,t) & R_{Bp22}(L,t) \end{pmatrix}\begin{pmatrix} \Theta(0^+) \\ b_p(0,0^+) \end{pmatrix}$$

(31)

where $R_{Bpij}$ stand for the coefficients of the matrix $\overline{R_{Bp}}(z,t)$. Then, the calculation can be carried out as in Ref. [1] to give

$$b_p(0,t) = R_{Bp21}(0,t) - R_{Bp22}(0,t)\frac{R_{Bp21}(L,t)}{R_{Bp22}(L,t)}$$

(32)

Consequently using the definition of the reflection coefficient, one finds the indicial response $\hat{R}_\Theta$ in terms of reflectivity at the time $\tilde{t}$ after switching on abruptly a constant intensity source at $t = 0$, or in other words, when applying a Heaviside unit-step input$\Theta$; thus, it resultsfor the indicial response in terms of reflectance

$$\hat{R}_\Theta(\tilde{t}) \equiv \left|b_p(0,t)\right]_0^{\tilde{t}}\Big|^2$$

(33)

From Eq.(31) one also finds that

$$f_0(L,t) = \frac{Det[\overline{R_{Bp}}(L,t)]}{R_{Bp22}(L,t)}$$

(34)

The indicial response in terms of transmittance is

$$\hat{T}_\Theta(\tilde{t}) \equiv \left|f_0(L,t)\right]_0^{\tilde{t}}\Big|^2$$

(35)

### 3.3 Tagaki-Taupin approach

We now consider the particular case where $\tilde{\alpha}_B \approx \alpha_B$; this situation occurs when the grating is used in specular condition $p = 0$ (in this case $\tilde{\alpha}_B$ and $\alpha_B$ are strickly equal) or when the period of the grating is large with respect to the wavelength of the diffracted radiation so that $\kappa_p \approx \kappa_0$ and $\theta_p \approx \theta_0$. In this condition, it is useful to follow an approach adopted by Tagaki and Taupin and now usually implemented in the dynamical theory of x-ray diffraction[16], since it leads to a formulation giving aphysical insight into the problem, as we show hereafter. As shown in Appendix 2, the problem can be reduced to the following hyperbolic second order PDE

$$\mathcal{L}_{(v,w)}\left[\tilde{f}_0(v,w)\,;\,\tilde{b}_p(v,w)\right] = 0$$

(36)

where $\mathcal{L}_{(v,w)}$ is the differential operator defined by

$$\mathcal{L}_{(v,w)} = \left(\frac{\partial^2}{\partial v \partial w} + \frac{\pi^2}{\Lambda^2}\right)$$

(37)

with $\Lambda^2$, the quantity related to the coupling constants $K_B^+, K_B^-$ by

$$\Lambda^2 = -\frac{\pi^2}{K_B^+ K_B^-}$$

$v, w$ are the following characteristic coordinates

$$v = \frac{1}{2}(c\, t\, \sin\theta - z)$$

(39a)

$$w = \frac{1}{2}(c\, t\, \sin\theta + z)$$

(39b)

and $\tilde{f}_0(v,w)$, $\tilde{b}_p(v,w)$ the reduced field amplitudes defined according to

$$f_0(v,w) = \exp(-i\, a\, c\, t\, \sin\theta)\tilde{f}_0(v,w)$$

(40a)

$$b_p(v,w) = \exp(-i\, a\, c\, t\, \sin\theta)\tilde{b}_p(v,w)$$

(40b)

where $a \equiv \alpha_B$. Provided that the media are not absorbing, using Eqs.(17), it follows that

$$\Lambda = \frac{2\pi \sin\theta}{k\, |u_j\, U_{j,-p}|}$$

(41)

One recognizes that the quantity $\Lambda$ is the extinction length of the dynamical theory of diffraction [16]. The PDE given by Eq.(36) can be solved for given boundary conditions by implementing different methods; the most common one is the Riemann's method that we summarize hereafter. For the sake of consistency, we present in Appendix 3, a brief mathematical development of the Riemann's method applied to our problem; the rigorous mathematical foundations of the method can be found in Ref. [17] and some details of the calculation are available in Ref. [8]. The application of Riemann's method requires an integration contour in the characteristic coordinate plane shown in Figure 2.
12

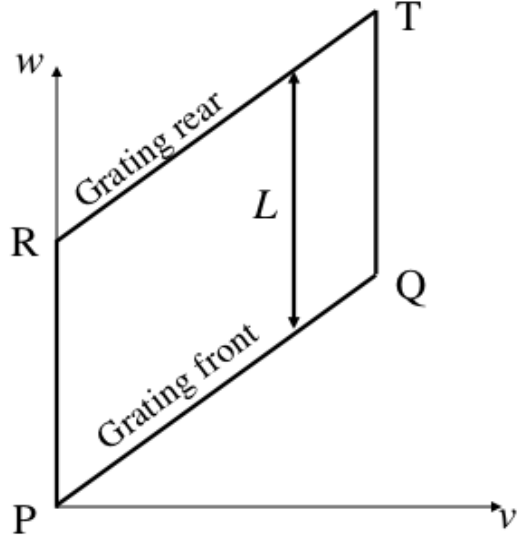

*Figure 2: Grating geometry in the characteristic coordinate reference frame (v,w); the grating front surfaceis given by w = v(line PQ) while the rear surface is given by w = v + L (line RT).*

As shown in Appendix 3,the diffracted field in reflection geometry can be written as

$$\tilde{b}_p(v,v) = -i\, K_B^- \frac{\Lambda}{\pi} \int_P^Q \frac{J_1\left[\frac{2\pi}{\Lambda}(v-v')\right]}{(v-v')} \tilde{f}_0(v',v')\, dv'$$

(42)

where $\tilde{f}_0(v',v')$ corresponds to the incoming wave at the front surface (z =0). $J_n$ is a Bessel function of the first kind. From Eq.(42) one can deduce the impulse and indicial response in terms of reflection coefficient. The impulse incident reduced field amplitude $\tilde{f}_0(v,v)_\delta$ reads

$$\tilde{f}_0(v,v)_\delta = \frac{\exp(+i\,2\,a\,v)\,\sin\theta}{\sqrt{2\pi}}\,\frac{\sin\theta}{2}\,\delta\left[\frac{x\cos\theta\,\sin\theta}{2} - v\right]$$

(43)

where $\delta$ stands for the Dirac peak. Inserting Eq.(43) in Eq.(42) and performing the integration gives for the reduced diffracted field $\tilde{b}_p(v,v)_\delta$ under the incidence of a Dirac pulse

$$\tilde{b}_p(v,v)_\delta = -i\,\sin\theta\, K_B^- \frac{\Lambda}{\pi}\,\frac{\exp(+i\,a\,x\cos\theta\,\sin\theta)}{\sqrt{2\pi}}\,\frac{J_1\left[\frac{\pi}{\Lambda}\sin\theta(c\,t - x\cos\theta)\right]}{\sin\theta(c\,t - x\cos\theta)}$$

(44)

that is, for the diffracted field

$$b_p(z=0,T)_\delta = i\sin\theta\, K_B^- \frac{\Lambda}{\pi\sqrt{2\pi}} \frac{J_1[\zeta(T)]}{\zeta(T)} \Theta(T)\,;\ \zeta(T) = \frac{\pi}{\Lambda}\sin\theta\, c\, T$$

(45)

where one has introduced the time delay $T$ measured with respect to the diffracted wave plane

$$T = \frac{c\,t - x\cos\theta}{c}$$

(46)

The quantity $g_R(T) = b_p(z=0,T)_\delta$ is the temporal Green function. For time coherent radiation with time-dependent causal distribution $\Xi$ (normalized to unity), the indicial response $\hat{R}_\Theta(t)$ in terms of reflection coefficient is given by

$$\hat{R}_\Theta(t) = \int_{-\infty}^{+\infty} |g_R(T)|^2\, \Xi(t-T)\, dT = \int_0^t |g_R(T)|^2\, \Xi(t-T)\, dT$$

(47)

Eq.(47) allows one to draw a "universal" curve for the indicial response $\hat{R}_\Theta(t)$ in terms of reduced time $\bar{t} = t\,\frac{\pi\sin\theta\, c}{\Lambda}$, see Figure 3.

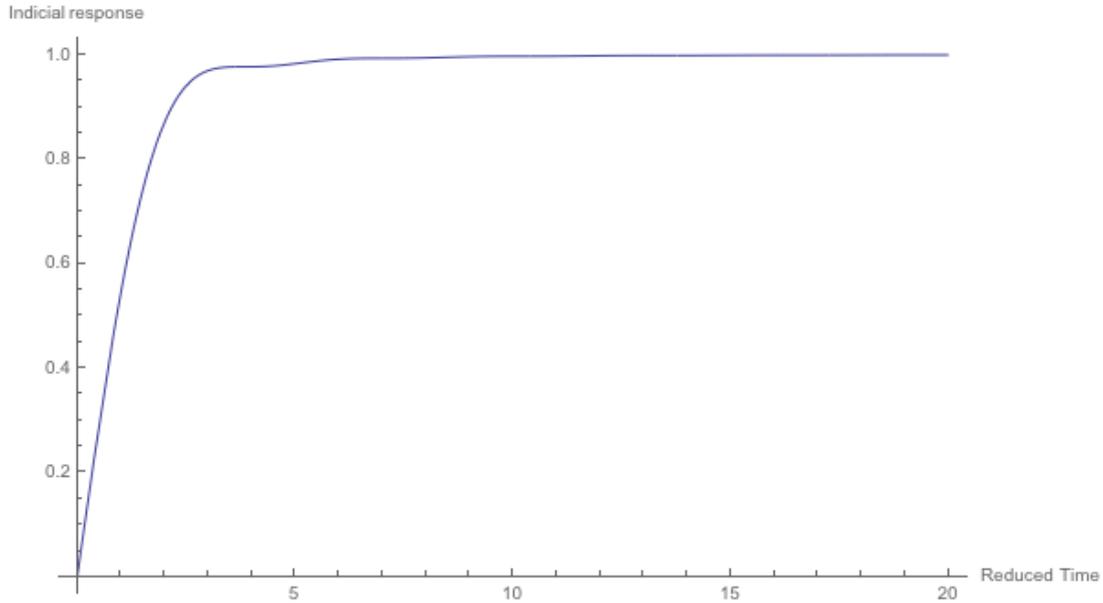

*Figure 3: Indicial response in terms of reflectance versus reduced time for a grating working in Bragg geometry close to the specular condition. The response is normalized to unity.*

It is interesting to note that the Tagaki-Taupin approach shows that the indicial response in terms of reflection is conditioned by the extinction length and that it presents a transient period whose duration is given by a characteristic transient time $t_c$ approximately equal to 2 units of reduced time as shown in figure 3. One is led to think that even where the specular approximation is not valid, the indicial response of the grating still presents a transient period of the order of $t_c$.

As also shown in Appendix 3, the diffracted field in transmission geometry is approximately given by :

$$\tilde{f}_0(v,w)\big]_T \approx -\frac{2\pi^2 L}{\Lambda^2} \int_P^Q \tilde{f}_0(v',v') \frac{J_1\left[\frac{2\pi}{\Lambda}\sqrt{(v+L-v')(v-v')}\right]}{\frac{2\pi}{\Lambda}\sqrt{(v+L-v')(v-v')}} dv'$$

(48)

To determine the impulse response in terms of transmission, we insert in Eq.(48) the expression of the incident pulse given by Eq.(44) as for the reflection case and we perform the integration as for the reflection geometry; it results that the impulse response in terms of transmission $f_0(z=L,T)_\delta$ is

$$f_0(z=L,T)_\delta \approx g_T(T) = \frac{\pi^2 L}{\sin\theta\, \Lambda^2} \frac{J_1[\xi(T)]}{\xi(T)} \Theta(T); \ \xi(T) = \frac{\pi}{\Lambda}\sqrt{cT\left(\frac{2L}{\sin\theta} + \frac{cT}{\sin\theta^2}\right)}$$

(49)

For time coherent radiation with time-dependent causal distribution $\Xi$ (normalized to unity), the indicial response $\hat{T}_\Theta(t)$ in terms of transmission coefficient is given by

$$\hat{T}_\Theta(t) = \int_{-\infty}^{+\infty} |g_T(T)|^2\, \Xi(t-T)\, dT = \int_0^t |g_T(T)|^2\, \Xi(t-T)\, dT$$

(50)

## 3.4 From indicial response to time-dependent reflection and transmission of a short pulse

Let $E(t)$ be the temporal envelope shape of any incident pulse and $S(t)$ the envelope of the time-dependent reflected or transmitted pulse. Then the Laplace transform $S(s)$ of $S(t)$ is related to the Laplace transform $E(s)$ of $E(t)$ by means of the convolution theorem

$$S(s) = \hat{Z}_\delta(s)\, E(s)$$

(51)

where $\hat{Z}_\delta(s)$ is the transfer function in terms of reflectance or transmittance that is the Laplace transform of the impulse response $\hat{Z}_\delta(t)$. This impulse response corresponds to the instantaneous reflectance or transmittance obtained when the system is struck by a Dirac pulse at *t=0* and should not be confused with the indicial response $\hat{Z}_\Theta(t)$. Nevertheless $\hat{Z}_\delta(t)$ and consequently $\hat{Z}_\delta(s)$ can be determined from $\hat{Z}_\Theta(t)$ by different methods; we have applied a basic but efficient graphic method introduced by Strejc[18,19]. The temporal dependence of the reflection or transmission of a short pulse can, then, be determined using Eq.(51) and then performing an inverse Laplace transform. Details can be found in Ref. [1].

## 4. Numerical simulations

We present a numerical example obtained with the time-dependent matrix approach in the continuation of the one presented in Ref. [1]. We consider a lamellar unslanted ($\phi = 0$) grating formed from multilayer bars. The multilayer structure consists of a periodic stack of N= 20Fe/C bilayers; the period $d$ is equal to 5.0 nm and the ratio $\gamma$ is equal to 0.5, that is to say the thicknesses of the Fe and C layers are the same. Figure 4 displays this structure and the parameters.

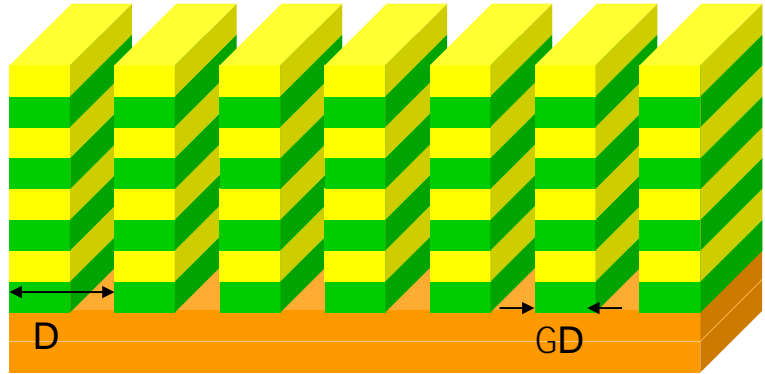

*Figure 4: Scheme of the unslanted lamellar grating considered in the numerical applications. D is the period of the grating and ΓD the width of the multilayer bars.*

The energy of the incident radiation is 8 keV. In all calculations, we use for the optical indices the values tabulated in the CXRO database [20]. One assumes that the grating period D equals 1 µm. Figures 5 and 6 show the steady state reflectance and transmittance respectively, for the diffraction order $p = $ -1 at different values of the Γ parameter: 0.1, 0.2, 0.4 and 0.6; they are calculated by the TWT presented in section 3.1 Let us emphasize that in the Bragg domain, the results of the TWT are in very good agreement with results given by a rigorous coupled wave analysis RCWA (see for instance Eq.(12) of Ref. [12]) or a modal theory (MT) [21] as illustrated by Figure 7. In this figure, as an example of the accuracy of the TWT in the Bragg domain, the steady-state reflectance for the unslanted grating with a Γ value of 0.2 is presented: the data from TWT are given by the solid line and the data from RCWA and MT (which are identical) by the dots; the calculations in the RCWA or MT are done with 15 Fourier terms(or modes). Similar agreements are observed for the other reflectances and transmittances displayed in figures 5 and 6.

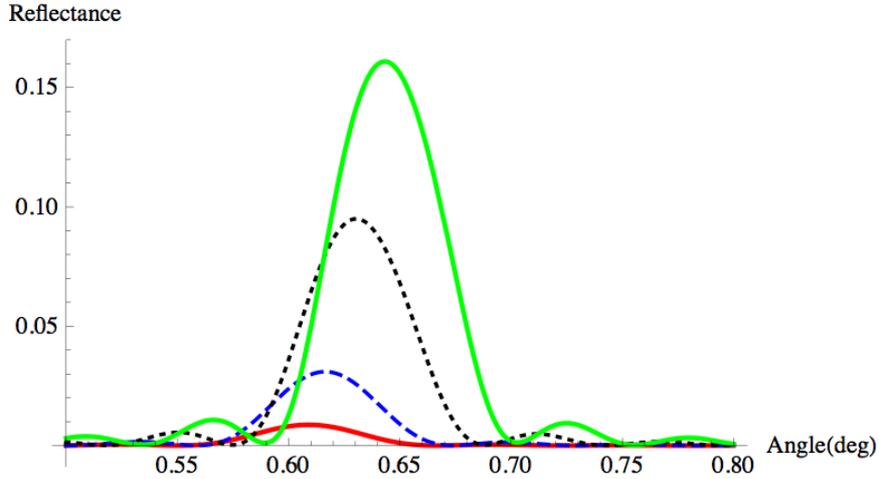

*Figure 5: Steady-state reflectance versus glancing angle $\theta_0$ of an unslanted lamellar multilayer grating shown in figure 4 for different values of the parameter $\Gamma$. The photon energy is 8 keV and the parameters of the grating are given in the text: $\Gamma=0.6$ green thick solid line, $\Gamma=0.4$ black dotted line, $\Gamma=0.2$ blue dashed line, $\Gamma=0.1$ red thick solid line. The diffraction order $p = -1$.*

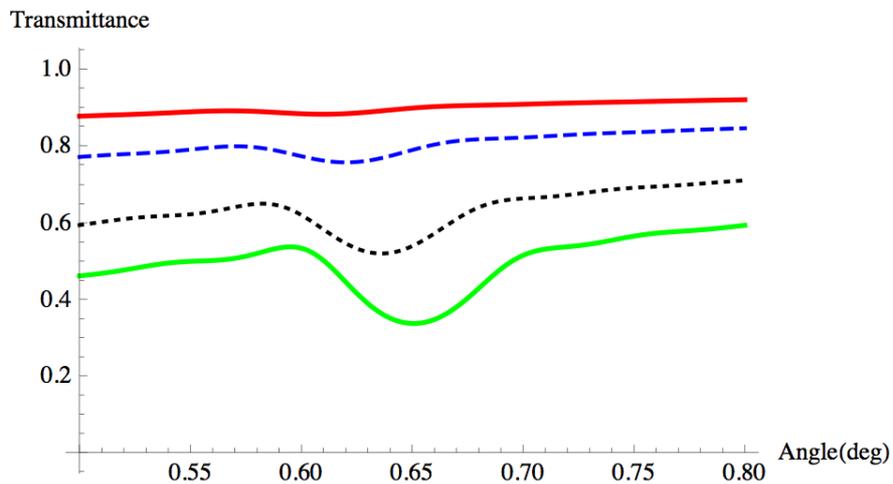

*Figure6: Steady-state transmittance versus glancing angle $\theta_0$ of an unslanted lamellar multilayer grating shown in figure 4 for different values of the parameter $\Gamma$. The photon energy is 8 keV and the parameters of the grating are given in the text: $\Gamma=0.6$ green thick solid line, $\Gamma=0.4$ black dotted line, $\Gamma=0.2$ blue dashed line, $\Gamma=0.1$ red thick solid line. The diffraction order $p = -1$.*

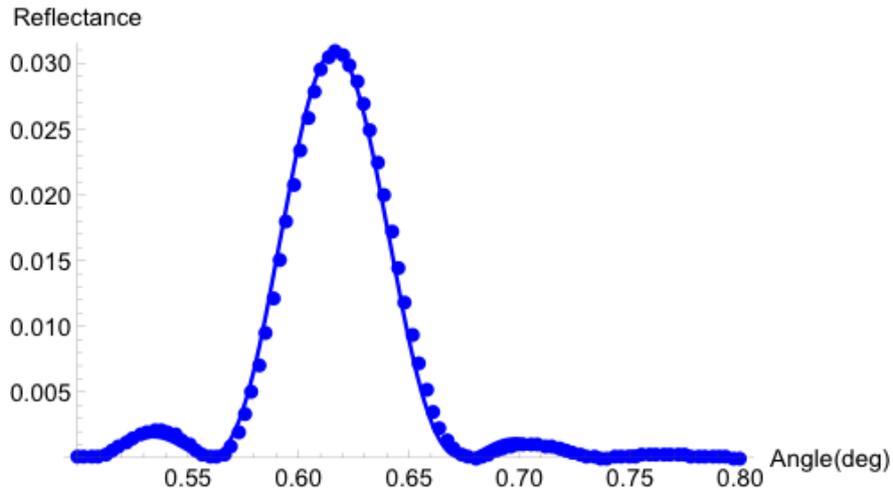

*Figure 7: Steady-state reflectance versus glancing angle $\theta_0$ of the unslanted lamellar multilayer grating shown in figure 4 for $\Gamma=0.2$ calculated from TWT (solid line) and rigorous theories RCWA and MT (dots). The diffraction order $p = -1$.*

Figure 8 shows the indicial response of the peak reflectance $\hat{R}_{\theta_{Bragg}}$ for the grating diffraction order $p = -1$, that is the reflectance at the generalized Bragg angle $\theta_{Bragg}$, for different values of the $\Gamma$ parameter: 0.1, 0.2, 0.4 and 0.6. It appears that the response presents a transient period in agreement with the Tagaki-Taupin approach; the transient time is approximately equal to two units of reduced time and depends on the value of the $\Gamma$ parameter: the smaller the $\Gamma$, the smaller the $\Lambda$ (Eq.(42)) since $U_{j,-p}$ is proportional to $\Gamma$ (Eq.(4)) and consequently, the longer the transient period as indicated by the Tagaki-Taupin theory. Let us outline that the generalized Bragg angle varies slightly with the value of the $\Gamma$ parameter according to Eq.(14). After the transient period the peak reflectance reaches the steady state value $R_{sp}$ which also depends on the value of $\Gamma$ [12]. Figure 9 shows the indicial response $\hat{T}_{\theta_{Bragg}}$ for the grating diffraction order $p = -1$, now in terms of transmittance at the generalized Bragg angle $\theta_{Bragg}$, for different values of the $\Gamma$ parameter: 0.1, 0.2, 0.4 and 0.6. The indicial response in terms of transmittance presents also a transient period but contrary to the reflectance case, the transient time depends not only on the $\Lambda$ but also on the thickness of the grating.

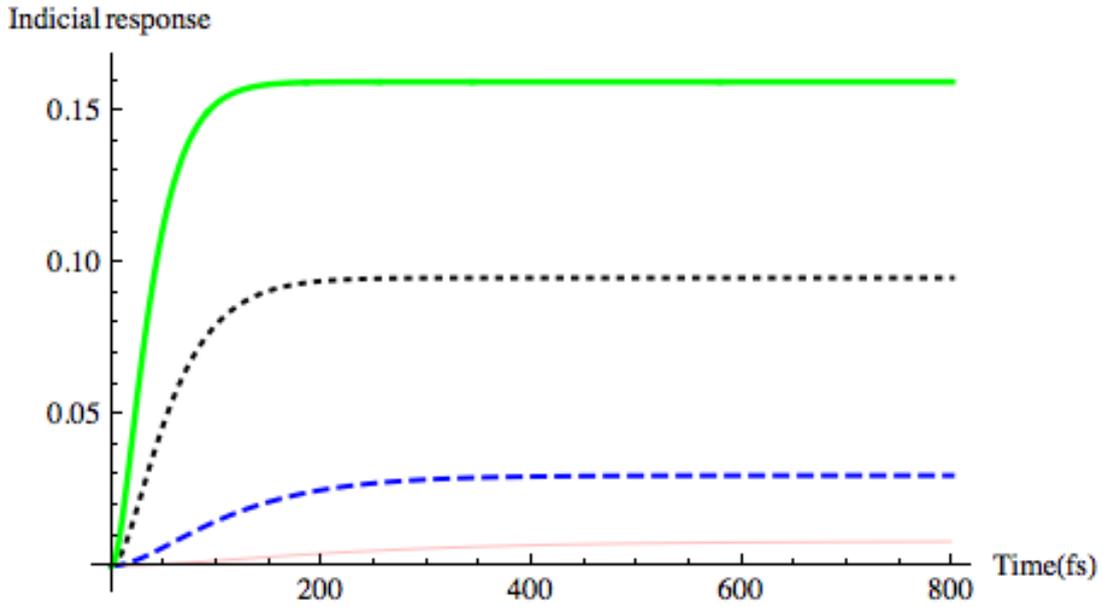

*Figure 8: Indicial response in terms of peak reflectance $\hat{R}_{\theta_{Bragg}}$ of an unslanted lamellar multilayer grating shown in figure4 for different values of the parameter $\Gamma$. The photon energy is 8 keV, the grating diffraction order p = -1 and the parameters of the grating are given in the text: $\Gamma$=0.6 green thick solid line, $\Gamma$=0.4 black dotted line, $\Gamma$=0.2 blue dashed line, $\Gamma$=0.1 red thin solid line.*

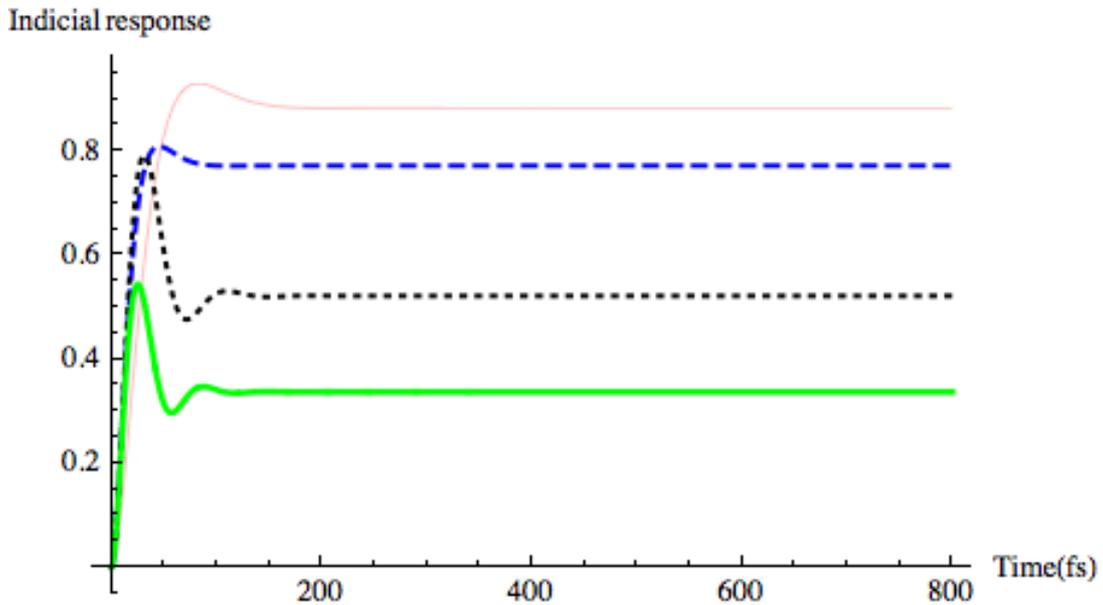

*Figure 9: Indicial response in terms of transmittance $\hat{T}_{\theta_{Bragg}}$ of an unslanted lamellar multilayer grating shown in figure 4 for different values of the parameter $\Gamma$ at the Bragg angle. The photon energy is 8 keV, the grating diffraction order p = -1 and the parameters of the grating are given in the text: $\Gamma$=0.6 green thick solid line, $\Gamma$=0.4 black dotted line, $\Gamma$=0.2 blue dashed line, $\Gamma$=0.1 red thin solid line.*

Figures 10 and 11 display the 3-dimensional temporal and spectral indicial response in terms of reflectance and transmittance for the unslanted lamellar grating described in figure 4 for $\Gamma=0.2$. The spectral dependence versus the glancing angle at a given photon energy (8keV) is given; we choose this representation in order to avoid taking into account the energy dispersion of the optical indices. The figures are zoomed on the short values of time in order to focus on the transient period. It clearly appears that the transient time is shorter for transmittance than for reflectance; as mentioned previously the transient time in transmittance is mainly dependent on the thickness $L$ of the grating, in agreement with the Tagaki-Taupin theory (see Eq.(48)).

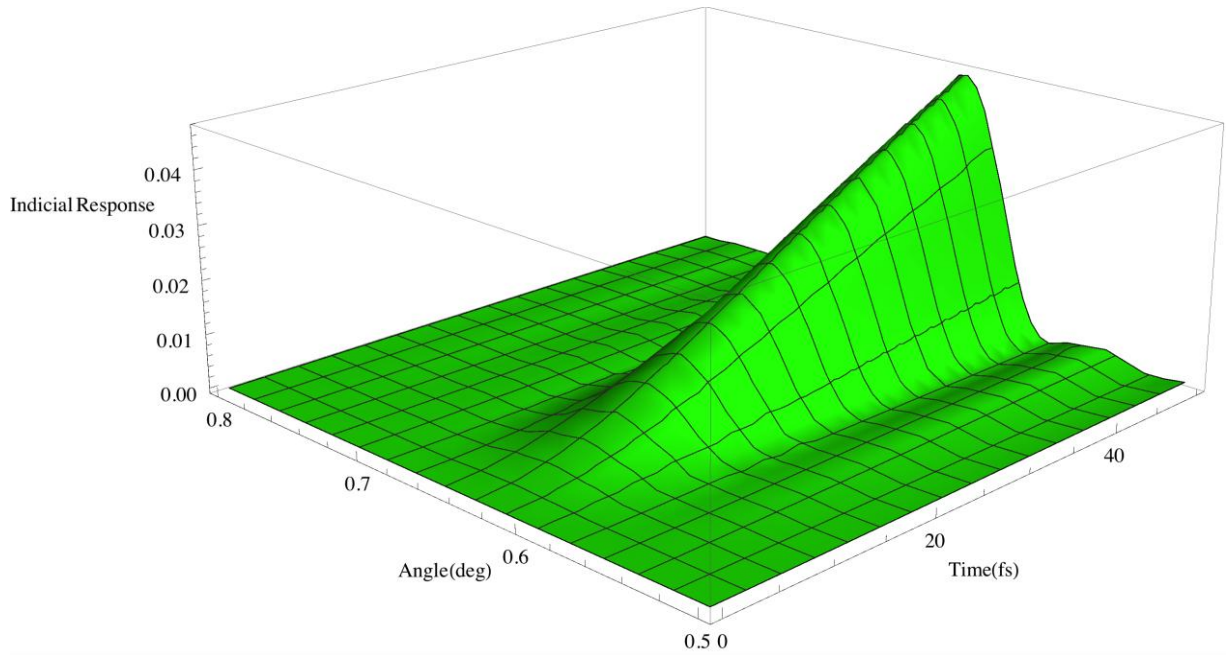

*Figure 10: Temporal and spectral indicial response in terms of reflectance of the lamellar grating shown in figure 4 for $\Gamma=0.2$. The photon energy is 8 keV, the grating diffraction order p = -1.*

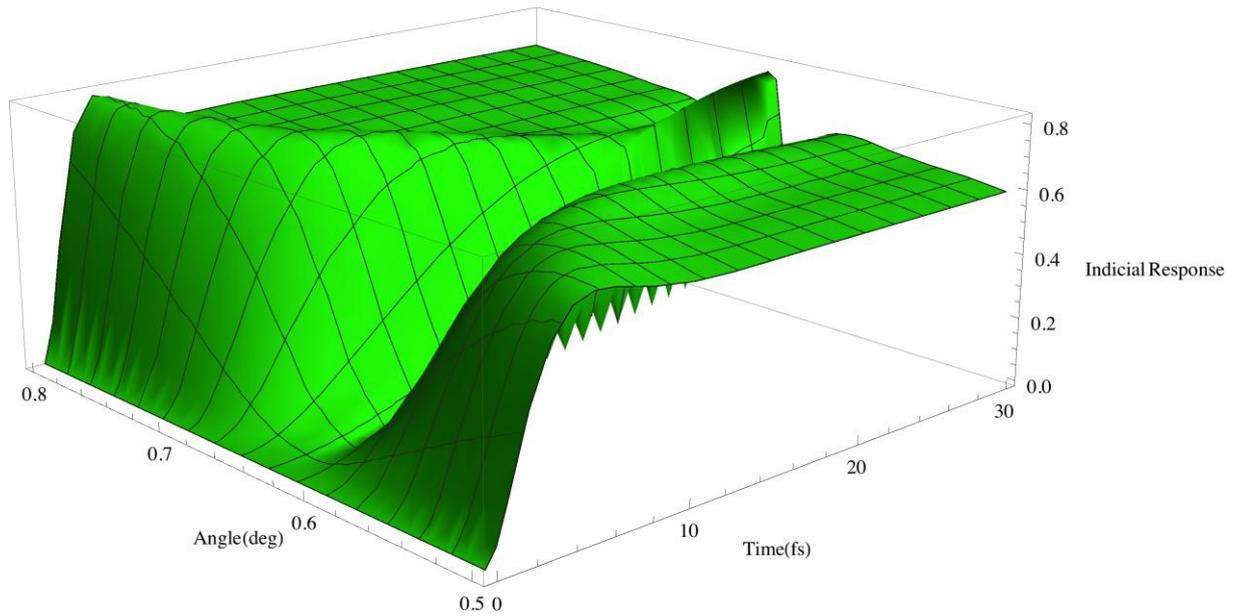

*Figure 11: Temporal and spectral indicial response in terms of transmittance of the lamellar grating shown in figure 4 for Γ=0.2. The photon energy is 8 keV, the grating diffraction order p = -1.*

Figure 12 displays the reflected peak height of an incident Gaussian pulse with unit amplitude and temporal width equal to 10 fs for the grating considered in figure 4. The calculations are performed according to the method given in section 3.4. The influence of the Γ parameter is clearly evident, particularly concerning the stretching of the reflected pulse. The calculations are carried out according to the method given in section 3.4.

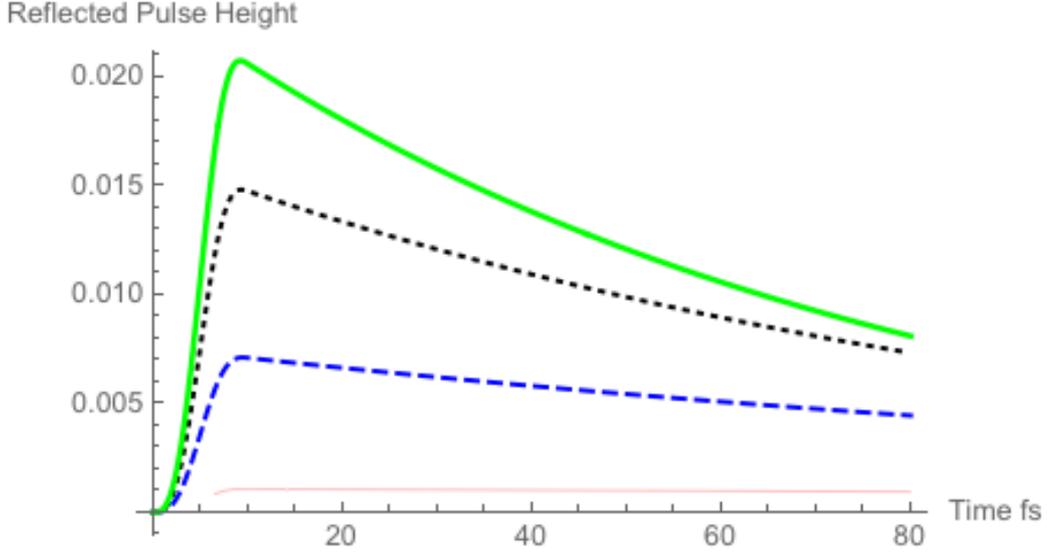

*Figure 12: Time dependence of the reflected pulse height for an incident pulse of width equal to 10 fs, of an unslanted lamellar multilayer grating shown in figure 4 for different values of the parameter Γ. The photon energy is 8 keV, the grating diffraction order p = -1 and the parameters of the grating are given in the text: Γ=0.6 green thick solid line, Γ=0.4 black dotted line, Γ=0.2 blue dashed line, Γ=0.1 red thin solid line.*

## 5. Conclusion

We have generalized the time-dependent coupled-wave theory initially developed for one-dimensional photonic crystals [1] to the spatiotemporal diffraction by multilayer gratings. The results obtained with the matrix formalism are in agreement with the Tagaki-Taupin theory originally developed in the framework of the dynamical theory of crystal diffraction that we have extended to grating diffraction. From Tagaki-Taupin theory, it appears that in reflection geometry, the key quantity is the extinction length while in transmission geometry, the thickness of the grating is also an important parameter. Although a very interesting means for obtaining a physical insight of the problem, Tagaki-Taupin approach is not relevant for predicting accurately the non-specular case behaviour; instead a matrix approach is very efficient.

This theoretical work gives a useful tool to predict the temporal response of optics implemented with short sources such as free electron lasers, high harmonic generation sources, …Numericalapplications of the present theory to gratings other than the lamellar one will be presented in a forthcoming paper. Our time-dependent matrix approach should be applicable beyond the two-wave approximation that is in the framework of the rigorous coupled-wave analysis; this work is underway. Finally this theory has to be extended to *p*-polarizationand conical diffraction that requires going beyond the paraxial approximation.

# APPENDIX 1

Inserting Eq.(20) in Eq.(15) leads to eigenvalue problem

$$(\overline{\mathcal{M}_p} - \psi \bar{I}) \begin{pmatrix} A \\ B \end{pmatrix} = 0$$

(A1.1)

Solving this problem gives the eigenvalues $\psi^{\pm}$ of $\overline{\mathcal{M}_p}$

$$\psi^{\pm} = \frac{(\widetilde{\alpha_B} - \alpha_B) \pm q}{2} \; ; q = \sqrt{4 K_B^+ K_B^- + (\widetilde{\alpha_B} + \alpha_B)^2}$$

(A1.2)

To the two eigenvalues $\psi^{\pm}$ are associated two eigenvectors $\bar{V}^{\pm}$

$$\bar{V}^{\pm} = \begin{pmatrix} \frac{-(\widetilde{\alpha_B} + \alpha_B) \mp q}{2 K_B^-} \\ 1 \end{pmatrix}$$

(A1.3)

The solutions for $\overline{\Sigma_B}(z)$ can be derived using the eigenmatrix

$$\bar{P} = (\bar{V}^+ \quad \bar{V}^-)$$

(A1.4)

that is

$$\overline{\Sigma_B}(z) = \bar{P} \begin{pmatrix} e^{+i\psi^+ z} & 0 \\ 0 & e^{i\psi^- z} \end{pmatrix} \bar{P}^{-1} \overline{\Sigma_B}(0)$$

(A1.5)

Since the general solution is a linear combination of the eigensolutions

$$\overline{\Sigma_B}(z) = C^+ e^{+i\psi^+ z} \bar{V}^+ + C^- e^{i\psi^- z} \bar{V}^- = \bar{P} \begin{pmatrix} e^{+i\psi^+ z} & 0 \\ 0 & e^{i\psi^- z} \end{pmatrix} \begin{pmatrix} C^+ \\ C^- \end{pmatrix}$$

(A1.6)

At $z = 0$, one has

$$\begin{pmatrix} C^+ \\ C^- \end{pmatrix} = \bar{P}^{-1} \overline{\Sigma_B}(0)$$

(A1.7)

Putting Eq.(A1.7) in Eq (A1.5), it follows that

$$\overline{\Sigma_B}(z) = \overline{S_{Bp}}(z) \overline{\Sigma_B}(0)$$

(A1.8)

where $\overline{S_{Bp}}(z)$ is obtained by the product $\bar{P} \begin{pmatrix} e^{+i\psi^+ z} & 0 \\ 0 & e^{i\psi^- z} \end{pmatrix} \bar{P}^{-1}$.

# APPENDIX 2

With $\tilde{\alpha}_B \approx \alpha_B \equiv a$ and $\theta_p \approx \theta_0 \equiv \theta$, the system given by Eq.(15) can be written

$$\left(sin\theta \frac{\partial}{\partial z} + \frac{1}{c}\frac{\partial}{\partial t} + i\, a\, sin\theta\right) f_0(z,t) = i\, sin\theta\, K_B^+ b_p(z,t)$$

(A2.1)

$$\left(-sin\theta \frac{\partial}{\partial z} + \frac{1}{c}\frac{\partial}{\partial t} + i\, a\, sin\theta\right) b_p(z,t) = -i\, sin\theta\, K_B^- f_0(z,t)$$

(A2.2)

The system given by Eqs.(A2.1,A2.2) can be rewritten in terms of the characteristic coordinates Eqs.(39) and reduced fields Eqs.(40) as

$$\frac{\partial}{\partial w}\tilde{f}_0(v,w) = i\, K_B^+ \tilde{b}_p(v,w)$$

(A2.3)

$$\frac{\partial}{\partial v}\tilde{b}_p(v,w) = -i\, K_B^- \tilde{f}_0(v,w)$$

(A2.4)

This new system of coupled PDEs is formally the same as the system of Tagaki-Taupin equations of the dynamical theory of x-ray diffraction and can be solved by using the same mathematical techniques. To do it, we first perform for each equation of the system a second differentiation with respect to the second characteristic variable, which gives

$$\frac{\partial^2}{\partial v \partial w}\tilde{f}_0(v,w) = i\, K_B^+ \frac{\partial}{\partial v}\tilde{b}_p(v,w)$$

(A2.5)

and

$$\frac{\partial^2}{\partial v \partial w}\tilde{b}_p(v,w) = -i\, K_B^- \frac{\partial}{\partial w}\tilde{f}_0(v,w)$$

(A2.6)

Combining Eqs.(A2.3-A2.6) results in the following hyperbolic second order PDE

$$\left(\frac{\partial^2}{\partial v \partial w} - K_B^+ K_B^-\right)[\tilde{f}_0(v,w)\, ;\, \tilde{b}_p(v,w)] = 0$$

(A2.7)

# APPENDIX 3

Let $R(v,w;v',w')$ be the Riemann function which satisfies

$$\mathcal{L}_{(v',w')}[R(v,w;v',w')] = 0$$

(A3.1)

To treat the Bragg-case geometry, we consider the contour $\mathcal{C}$ formed by the triangle *(PQR)* formed in the plane of characteristic coordinates $v, w$ as shown in Figure 2; the segment $\overline{PQ}$ corresponds mathematically to the condition $v = w$ and physically to the front surface of the grating ($v = w \Rightarrow z = 0$); $\Delta$ is the surface domain enclosed by $\mathcal{C}$. If $F(v, w)$ is solution of

$$\mathcal{L}_{(v,w)}[F(v,w)] = 0$$

(A3.2)

then taking into account Eqs.(A3.1, A3.2), the surface integral cancels.

$$I_\Delta = \iint_\Delta \left( R\, \mathcal{L}_{(v,w)}[F] - F\, \mathcal{L}_{(v,w)}[R] \right) dv\, dw$$

(A3.3)

Applying Green's identity, $I_\Delta$ can be transformed in contour integral

$$I_\Delta = \oint_\mathcal{C} (\widehat{\boldsymbol{n}} \cdot \boldsymbol{D})\, dl = 0$$

(A3.4)

with $\widehat{\boldsymbol{n}}$ unit vector outward pointing normal to the line element $dl$ and

$$\boldsymbol{D} = \frac{1}{2}\left[\left(R\,\frac{\partial F}{\partial v} - F\,\frac{\partial R}{\partial v}\right)\widehat{\boldsymbol{v}} + \left(R\,\frac{\partial F}{\partial w} - F\,\frac{\partial R}{\partial w}\right)\widehat{\boldsymbol{w}}\right]$$

(A3.5)

If the field $F(v, w)$ is assumed to vanish along the line $\overline{PR}$, integration by parts leads to

$$R(v,w;v',w')F(v',w')]_Q - \int_P^Q F(v',w')\frac{\partial R(v,w;v',w')}{\partial v'}\,dv'$$
$$- \int_Q^P \left(F(v',w')\frac{\partial R(v,w;v',w')}{\partial w'} + R(v,w;v',w')\frac{\partial F(v',w')}{\partial v'}\right) dl = 0$$

(A3.6)

In addition one requires that the three following conditions are satisfied:

$$\frac{\partial R(v,w;v',w')}{\partial w'} = 0$$

(A3.7)

along the line $\overline{PQ}\ (v' = w')$,

$$\frac{\partial R(v,w;v',w')}{\partial v'} = 0$$

(A3.8)

along the line $\overline{RQ}$ ($w' = w$), and also at the point $Q$

$$R(v, w; v', w')]_Q = 1$$

(A3.9)

Introducing these conditions in Eq.(A3.6) gives

$$F(v', w')]_Q = F(v', v') = \int_Q^P R(v, w; v', w') \frac{\partial F(v', w')}{\partial v'} dl$$

(A3.10)

Let $F(v', w')$ be $\tilde{b}_p(v', w')$, then from Eq.(A2.4), $\frac{\partial F(v', w')}{\partial v'} = -i K_B^- \tilde{f}_0(v, w)$ and from Eq.(A3.10)

$$\tilde{b}_p(v', w')]_Q = \tilde{b}_p(v', v') = -i K_B^- \int_P^Q R(v, w; v', w') \tilde{f}_0(v', v')]_Q dv'$$

(A3.11)

Afanas'ev and Kohn [22]in the context of a Tagaki-Taupin problem of diffraction have given the Riemann $R(v, w; v', w')$ function that fulfills the reflection conditions given by Eqs.(A3.7-A3.9):

$$R(v, w; v', w') = J_0\left[\frac{2\pi}{\Lambda}\sqrt{(w-w')(v-v')}\right] + \frac{w-w'}{v-v'} J_2\left[\frac{2\pi}{\Lambda}\sqrt{(w-w')(v-v')}\right]$$

(A3.12)

Since $v'$ is equal to $w'$ along the line $\overline{PQ}$, the Riemann function $R(v, v; v', v')$ reduces to

$$R(v, v; v', v') = J_0\left[\frac{2\pi}{\Lambda}(v-v')\right] + J_2\left[\frac{2\pi}{\Lambda}(v-v')\right] = \frac{\Lambda J_1\left[\frac{2\pi}{\Lambda}(v-v')\right]}{\pi(v-v')}$$

(A3.13)

where the following relationship has been used

$$J_0[x] + J_2[x] = \frac{2 J_1[x]}{x}$$

(A3.14)

Finally combining Eqs.(A3.11,A3.13), the reflected field can be written

$$\tilde{b}_p(v, v) = -i K_B^- \frac{\Lambda}{\pi} \int_P^Q \frac{J_1\left[\frac{2\pi}{\Lambda}(v-v')\right]}{(v-v')} \tilde{f}_0(v', v') dv'$$

(A3.15)

To treat the transmission geometry, we consider the contour $\mathcal{B}$ formed by the parallelogram *(PRTQ)* formed in the plane of characteristic coordinates *v*, *w*,as shown in Figure 2; the segment $\overline{RT}$ corresponds mathematically to the condition $v = w-L$ and physically to the rear surface of the grating. In a way similar to the reflection case, contour integration along $\mathcal{B}$ leads to

$$R(v,w;v',w')F(v',w')]_T - \int_T^Q F(v',w')\frac{\partial R(v,w;v',w')}{\partial w'} dw' +$$

$$\int_Q^P (F(v',w')\frac{\partial R(v,w;v',w')}{\partial w'} + R(v,w;v',w')\frac{\partial F(v',w')}{\partial v'}) dl$$

$$- \int_R^T (F(v',w')\frac{\partial R(v,w;v',w')}{\partial v'} + R(v,w;v',w')\frac{\partial F(v',w')}{\partial w'}) dl = 0$$

(A3.16)

Following additional conditions (similar to the reflection case) are required

$$\frac{\partial R(v,w;v',w')}{\partial w'} = 0$$

(A3.17)

along the line $\overline{TQ}$

$$\frac{\partial R(v,w;v',w')}{\partial v'} = 0$$

(A3.18)

along the line $\overline{RT}$, and also at the point $T$

$$R(v,w;v',w')]_T = 1$$

(A3.19)

Implementing these conditions Eqs.(A3.17-A3.19) in Eq.(A3.16) gives

$$F(v',w')]_T = R(v,w;v',w')F(v',w')]_Q - \int_P^Q R(v,w;v',w')\frac{\partial F(v',w')}{\partial v'} dv'$$

$$- \int_P^Q F(v',w')\frac{\partial R(v,w;v',w')}{\partial v'}[w'=v' dv'$$

(A3.20)

Afanas'ev and Kohn [21] have given the Riemann $R(v,w;v',w')$ function that fulfills the transmission conditions given by Eqs.(A3.17-A3.19)

$$R(v,w;v',w') = J_0\left[\frac{2\pi}{\Lambda}\sqrt{(w-w')(v-v')}\right] + \frac{v-v'}{w-w'}J_2\left[\frac{2\pi}{\Lambda}\sqrt{(w-w')(v-v')}\right]$$

(A3.21)

along the rear surface $w = v + L$ and

$$\frac{\partial R(v,w;v',w')}{\partial v'} = \frac{2\pi^2 L}{\Lambda^2} \frac{J_1\left[\frac{2\pi}{\Lambda}\sqrt{(v+L-v')(v-v')}\right]}{\frac{2\pi}{\Lambda}\sqrt{(v+L-v')(v-v')}}$$

(A3.22)

Let's $F(v,w)$ be $\tilde{f}_0(v,w)$, then from Eq.(A2.3), $\frac{\partial F(v',w')}{\partial v'} = \frac{\partial}{\partial w'}\tilde{f}_0(v',w') = i\, K_B^+ \tilde{b}_p(v',w')$. Then

$$\tilde{f}_0(v,w)]_T = R(v,w;v',w')\tilde{f}_0(v',w')]_Q - i\, K_B^+ \int_P^Q R(v,w;v',w')\tilde{b}_p(v',w')\, dv'$$

$$- \int_P^Q \tilde{f}_0(v',v')\frac{\partial R(v,w;v',w')}{\partial v'}[w'=v'\, dv'$$

(A3.23)

The first term of Eq.(A3.23) is the initial conditions and will be omitted; the last term dominates the second term since it is generated by the incident($\tilde{f}_0$) and not diffracted pulse($\tilde{b}_p$); finally one has

$$\tilde{f}_0(v,w)]_T \approx -\frac{2\pi^2 L}{\Lambda^2}\int_P^Q \tilde{f}_0(v',v')\frac{J_1\left[\frac{2\pi}{\Lambda}\sqrt{(v+L-v')(v-v')}\right]}{\frac{2\pi}{\Lambda}\sqrt{(v+L-v')(v-v')}}\, dv'$$

(A3.24)